\documentclass[copyright,creativecommons]{eptcs}
\providecommand{\event}{DCM'09} \providecommand{\volume}{??} \providecommand{\anno}{20??}
\providecommand{\firstpage}{1}
\usepackage{breakurl}
\usepackage{amsmath, amsfonts, amssymb, eucal, amsthm}

\input{Qcircuit}

\newcommand{\quotes}[1]{``#1''}
\newcommand{\comment}[1]{}
\newcommand{\braket}[2]{{\langle {#1}\!\mid\!{#2} \rangle}}
\newcommand{\Hilbert}{{\cal H}}
\newcommand{\GoodSetSize}[1]{2^{\lceil\log((2/\epsilon)\ln{2#1})\rceil}}
\newcommand{\hsp}[1]{\ensuremath{\text{HSP}_{G,K}\left(#1\right)}}

\newtheorem{definition}{Definition}%[section]
\newtheorem{lemma}{Lemma}%[section]
\newtheorem{theorem}{Theorem}%[section]

\begin{document}

\title{Algorithms for Quantum Branching Programs\\ Based on Fingerprinting}
\author{Farid Ablayev\thanks{Work was in part supported by the Russian Foundation for Basic Research
    under the grant ¹08-07-00449a}
\institute{Institute for Informatics\\ Kazan, Russian Federation}
\email{fablayev@gmail.com} \and Alexander Vasiliev
\thanks{Work was in part supported by
the Russian Foundation for Basic Research
    under the grant ¹08-07-00449a} \institute{Institute
for Informatics\\ Kazan, Russian Federation} \email{Alexander.Vasiliev@ksu.ru} }
\def\titlerunning{Algorithms for Quantum Branching Programs}
\def\authorrunning{F. Ablayev \& A. Vasiliev}

%\institute{Dept. of Theoretical Cybernetics, \\ Kazan State
%University\\
%420008 Kazan, Russia, \\
%\email{ablayev@ksu.ru, Ayrat.Khasyanov@ksu.ru, vaslo@mail.ru}\\
%}

\maketitle

\begin{abstract}
In the paper we develop a  method for constructing quantum
algorithms for computing Boolean functions by quantum ordered
read-once branching programs (quantum OBDDs). Our method is based on
fingerprinting technique and representation of Boolean functions by
their characteristic polynomials. We use circuit notation for
branching programs for desired algorithms presentation. For several
known functions our approach provides optimal
QOBDDs. %and which are cheaper than their classical stochastic
%counterparts.
Namely we consider such functions as $MOD_m$, $EQ_n$,
$Palindrome_n$, and $PERM_n$ (testing whether given Boolean matrix
is the Permutation Matrix). We also propose a generalization of our
method and apply it to the Boolean variant of the \emph{Hidden
Subgroup Problem}.
\end{abstract}

%%%%%%%%%%%%%%%%%%%%%%%%%%%%%%%%%%%%%%%%%%%%%%%%%%%%%%

\section{Introduction}

During the last two decades different types of quantum computation
models based on Turing Machines,  automata, and  circuits have been
considered.  For some of them different examples of functions were
presented for which quantum models appear to be much more
(exponentially) efficient than their classical counterparts.

In this paper we consider a restricted model of computation known as
\emph{Ordered Read-Once Quantum Branching Programs}. In computer
science this model is also known as Ordered Binary Decision Diagrams
(OBDDs). The main reason for the investigation of restricted models
of quantum computers was proposed by Ambainis and Freivalds in 1998
\cite{af98}. Considering one-way quantum finite automata, they
suggested that first quantum-mechanical computers would consist of a
comparatively simple and fast quantum-mechanical part connected to a
classical computer.

Two models of \emph{quantum branching programs} were introduced by
Ablayev, Gainutdinova, Karpinski \cite{agk01} (\emph{leveled
programs}), and by Nakanishi, Hamaguchi, Kashiwabara \cite{nhk00}
(\emph{non-leveled programs}). Later it was shown by Sauerhoff
\cite{ss04} that these two models are polynomially equivalent.

For this model we develop the \emph{fingerprinting} technique
introduced in \cite{av08}. The basic ideas of this approach are due
to Freivalds (e.g. see the book \cite{mr95}). It was later
successfully applied in the \emph{quantum automata} setting by
Ambainis and Freivalds in 1998 \cite{af98} (later improved in
\cite{an08}). Subsequently, the same technique was adapted for the
quantum branching programs by Ablayev, Gainutdinova and Karpinski in
2001 \cite{agk01}, and was later generalized in \cite{av08}.

For our technique we use the presentation of Boolean functions known
as \emph{characteristic polynomials}. Our definition of the
characteristic polynomial differs from that of \cite{alw98}, though
it uses similar ideas.

We display several known functions for which our method provides
optimal QOBDDs. Namely, these functions are $MOD_m$, $EQ_n$,
$Palindrome_n$, and $PERM_n$.

%%%%%%%%%%%%%%%%%%%%%%%%%%%%%%%%%%%%%%%%%%%%%%%%%%%%%%
\section{Preliminaries}

We use the notation $\ket{i}$ for the vector from $\Hilbert^d$,
which has a $1$ on the $i$-th position and $0$ elsewhere. Obviously,
the set of vectors $\ket{1}$,\ldots,$\ket{d}$ forms an orthonormal
basis in $\Hilbert^d$.

\begin{definition}
A Quantum Branching Program ${Q}$ over the Hilbert space
$\Hilbert^d$ is defined as
\[
Q=\langle T, \ket{\psi_0}, M_{accept}\rangle,
\]
where $T$ is a sequence of $l$ instructions: $T_j=\left(x_{i_j},
U_j(0),U_j(1)\right)$ is determined by the variable $x_{i_j}$ tested
on the step $j$, and $U_j(0)$, $U_j(1)$ are unitary transformations
in $\Hilbert^d$.

Vectors $\ket{\psi}\in \Hilbert^d$ are called states (state vectors)
of $Q$, $\ket{\psi_0}\in \Hilbert^d$ is the initial state of $Q$,
and $M_{accept}$ -- is a projector on the accepting subspace
$\Hilbert^{d}_{accept}$ (i.e. it is a diagonal zero-one projection
matrix, which determines the final projective measurement).

We define a computation of ${Q}$ on an input $\sigma = (\sigma_1,
\ldots, \sigma_n) \in \{0,1\}^n$  as follows:
\begin{enumerate}
\item A computation of ${Q}$ starts from the initial state
      $\ket{\psi_0}$;
\item The $j$-th instruction of $Q$ reads the input symbol $\sigma_{i_j}$ (the
value of $x_{i_j}$) and applies the transition matrix $U_j =
      U_j(\sigma_{i_j})$ to the current state $\ket{\psi}$ to
      obtain the state
      $\ket{\psi'}=U_j(\sigma_{i_j})\ket{\psi}$;
\item The final state is
\[ \ket{\psi_\sigma}= \left(\prod_{j=l}^1 U_j(\sigma_{i_j})\right)
\ket{\psi_0}\enspace . \]
\item After the $l$-th (last) step of quantum transformation $Q$ measures
its configuration $\ket{\psi_\sigma}$, and the input $\sigma$ is
accepted with probability
\[ Pr_{accept}(\sigma)=\braket{\psi_\sigma M^\dag_{accept}}{M_{accept}\psi_\sigma}=
||M_{accept}\ket{\psi_\sigma}||^2_2. \]
\end{enumerate}

\end{definition}

\paragraph{Circuit representation.}

A QBP can be viewed as a quantum circuit aided with an ability to read classical bits as
control variables for unitary operations. That is any quantum circuit is a QBP which
does not depend essentially on its classical inputs.
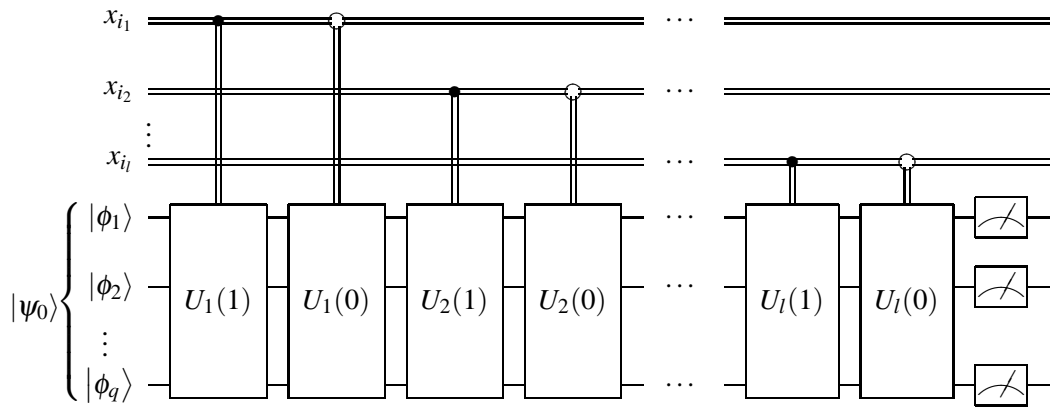
\[\quad\quad
\Qcircuit  @C=0.75em @R=1.0em {
&&&\lstick{x_{i_1}} & \control\cw\cwx[5] & \controlo\cw\cwx[5] & \cw & \cw & \cw & \push{\cdots\quad} & \cw & \cw & \cw & \cw\\
\\
&&&\lstick{x_{i_2}} & \cw & \cw & \control\cw\cwx[3] & \controlo\cw\cwx[3] & \cw & \push{\cdots\quad} & \cw & \cw & \cw & \cw\\
&&&\vdots \\
&&&\lstick{x_{i_l}} & \cw & \cw & \cw & \cw & \cw  & \push{\cdots\quad} & \control\cw\cwx[1] & \controlo\cw\cwx[1] & \cw & \cw\\
&&&\lstick{\ket{\phi_1}} & \multigate{3}{U_1(1)} & \multigate{3}{U_1(0)} & \multigate{3}{U_2(1)} & \multigate{3}{U_2(0)} & \qw  & \push{\cdots\quad} & \multigate{3}{U_l(1)} & \multigate{3}{U_l(0)} & \meter & \qw\\
&&&\lstick{\ket{\phi_2}} & \ghost{U_1(1)} & \ghost{U_1(0)} & \ghost{U_2(1)} & \ghost{U_2(0)} & \qw & \push{\cdots\quad} & \ghost{U_l(1)} & \ghost{U_l(0)} & \meter & \qw\\
\ustick{\ket{\psi_0}\quad\quad\quad~} \gategroup{6}{1}{9}{1}{1em}{\{} & \vdots \\
&&&\lstick{\ket{\phi_q}} & \ghost{U_1(1)} & \ghost{U_1(0)} & \ghost{U_2(1)} & \ghost{U_2(0)} & \qw & \push{\cdots\quad} & \ghost{U_l(1)} & \ghost{U_l(0)} & \meter & \qw\\
}
\]
Here $x_{i_1},\ldots,x_{i_l}$ is the sequence of (not necessarily
distinct) variables denoting classical control bits. Using the
common notation single wires carry quantum information and double
wires denote classical information and control.

\paragraph{Complexity measures.}
The \emph{width} of $Q$ is the dimension $d$ of the state space
$\Hilbert^d$, the \emph{length} of $Q$ is the number $l$ of
instructions in the sequence $T$.

Note that for a QBP in the circuit setting another important
complexity measure explicitly comes out -- a number $q$ of qubits
physically needed to implement a corresponding quantum system with
classical control. From definition it follows that $\log d\leq q$.

\begin{definition}
We call a quantum branching program a $q$-qubit QBP, if it can be
implemented as a classically-controlled quantum system based on $q$
qubits.
\end{definition}

\paragraph{Acceptance criteria.} A $\mathrm{QBP}$ $Q$ \emph{computes
the Boolean function $f$ with one-sided error} if there exists an
$\epsilon\in(0,1)$ (called an \emph{error}) such that for all
$\sigma\in f^{-1}(1)$ the probability of $Q$ accepting $\sigma$ is 1
and for all $\sigma\in f^{-1}(0)$ the probability of $Q$ erroneously
accepting $\sigma$ is less than $\epsilon$.

\paragraph{Read-once branching programs.}

\begin{definition}
We call a $\mathrm{QBP}$ $Q$ a quantum $\mathrm{OBDD}$
($\mathrm{QOBDD}$) or read-once $\mathrm{QBP}$ if each variable
$x\in\{x_1,\dots,x_n\}$ occurs in the sequence $T$ of
transformations of $Q$ at most once.
\end{definition}

For the rest of the paper we're only interested in QOBDDs, i.e. the
length of all programs would be $n$ (the number of input variables).
%The ``obliviousness'' is inherent for a QBP and therefore this
%definition is consistent with the usual notion of an OBDD.

\paragraph{Generalized Lower Bound.}
The following general lower bound on the width of QOBDDs was proven
in \cite{agkmp05}.

\begin{theorem}\label{quant-det-width}
 Let $f(x_1, \ldots, x_n)$ be a Boolean function
 computed by  a quantum read-once branching program $Q$. Then
 \[ {\rm width}(Q)=\Omega(\log {\rm width}(P)) \]
 where $P$  is a deterministic OBDD of minimal width computing
 $f(x_1, \ldots, x_n)$.
\end{theorem}

That is, the width of a quantum OBDD cannot be asymptotically less
than logarithm of the width of the minimal deterministic OBDD
computing the same function. And since the deterministic width of
many \quotes{natural} functions is exponential \cite{weg00}, we
obtain the linear lower bound for these functions.

\section{Algorithms for QBPs Based on Fingerprinting}

Generally \cite{mr95}, \emph{fingerprinting} -- is a technique that
allows to present objects (words over some finite alphabet) by their
\emph{fingerprints}, which are significantly smaller than the
originals. It is used in randomized and quantum algorithms to test
\emph{equality} of some objects (binary strings) with one-sided
error by simply comparing their fingerprints.

In this paper we develop a variant of the fingerprinting technique adapted for quantum
branching programs. At the heart of the method is the representation of Boolean
functions by polynomials of special type, which we call \emph{characteristic}.

\subsection{Characteristic Polynomials for Quantum Fingerprinting}

%\begin{definition}\label{polynomial-definition}
We call a polynomial $g(x_1,\dots,x_n)$ over the ring ${\mathbb
Z}_m$ a characteristic polynomial of a Boolean function
$f(x_1,\ldots,x_n)$ and denote it $g_f$
%
%if there exists number $m$ such that
%if $g(\sigma)=a \bmod{m}$ iff $f(\sigma)=1$.
when for all $\sigma\in\{0,1\}^n$ %for some $a\in {\mathbb Z}_m$ it holds
$g_f(\sigma)=0$ iff $f(\sigma)=1$.
%\end{definition}

%W.l.g. we assume  $a=0$ in Definition \ref{polynomial-definition}.
%
\begin{lemma}\label{Existence-Of-A-Characteristic-Polynomial}
For any Boolean function $f$ there exists a characteristic
polynomial $g_f$ over ${\mathbb Z}_{2^n}$.
\end{lemma}
%%%%%%%%%%%%%%%%%%%%%%%%%%%%%%%%%%%%%%%%%%%%
%{\em Proof.}
\begin{proof}
One way to construct such characteristic polynomial $g_f$ is
transforming a sum of products representation for $\neg f$.

Let $K_1\vee\ldots\vee K_l$ be a sum of products for $\neg f$ and
let $\tilde{K_i}$ be a product of terms from $K_i$ (negations $\neg
x_j$ are replaced by $1-x_j$). Then $\tilde{K_1}+\ldots+\tilde{K_l}$
is a characteristic polynomial over ${\mathbb Z}_{2^n}$ for $f$
since it equals $0$ $\iff$ all of $\tilde{K_i}$ (and thus $K_i$)
equal $0$.
This happens only when the negation of $f$ equals $0$. %\Endproof
\end{proof}

 Generally, there are many polynomials for the same function. For
example, the function $EQ_n$, which tests the equality of two
$n$-bit binary strings, has the following polynomial over ${\mathbb
Z}_{2^n}$:
$$\sum\limits_{i=1}^n\left(x_i(1-y_i)+(1-x_i)y_i\right)=\sum\limits_{i=1}^n\left(x_i+y_i-2x_iy_i\right).$$

On the other hand, the same function can be represented by the
polynomial
$$\sum\limits_{i=1}^nx_i2^{i-1}-\sum\limits_{i=1}^ny_i2^{i-1}.$$

We use this presentation of Boolean functions for our fingerprinting
technique which generalizes the algorithm for $MOD_m$ function by
Ambainis and Nahimovs \cite{an08}.

\subsection{Fingerprinting technique}\label{fqbp}

For a Boolean function $f$ we choose an error rate $\epsilon>0$ and
pick a characteristic polynomial $g$ over the ring $\mathbb{Z}_m$.
Then for arbitrary binary string $\sigma=\sigma_1\ldots \sigma_n$ we
create its fingerprint $\ket{h_\sigma}$ composing
$t=\GoodSetSize{m}$ single qubit fingerprints $\ket{h^i_\sigma}$:
$$\begin{array}{rcl}
\ket{h^i_\sigma} & = & \cos\frac{2\pi k_i
g(\sigma)}{m}\ket{0}+\sin\frac{2\pi k_i g(\sigma)}{m}\ket{1}\\
\ket{h_\sigma} & = &
\frac{1}{\sqrt{t}}\sum\limits_{i=1}^t\ket{i}\ket{h^i_\sigma}
\end{array}$$
That is, the last qubit is rotated by $t$ different angles about the
$\hat{y}$ axis of the Bloch sphere.

The chosen parameters $k_i\in\{1,\ldots,m-1\}$ for
$i\in\{1,\ldots,t\}$ are ``good'' following the notion of
\cite{af98}.

\begin{definition}\label{good-set}
A set of parameters $K=\{k_1, \dots, k_t\}$ is called ``good'' for
some integer $b\neq0 \bmod m$ if
$$\frac{1}{t^2}\left(\sum\limits_{i=1}^t\cos{\frac{2\pi k_i b}{m}}\right)^2<\epsilon.$$
\end{definition}
The left side of inequality is the squared amplitude of the basis
state $\ket{0}^{\otimes\log{t}}\ket{0}$ if $b=g(\sigma)$ and the
operator $H^{\otimes\log{t}}\otimes I$ has been applied to the
fingerprint $\ket{h_\sigma}$. Informally, that kind of set
guarantees, that the probability of error will be bounded by a
constant below 1.

The following lemma proves the existence of a ``good'' set and
generalizes the proof of the corresponding statement from
\cite{an08}.
%(for the proof see Appendix).

\begin{lemma}\cite{av08}\label{existence-of-a-good-set}
There is a set $K$ with $|K|=t=\GoodSetSize{m}$ which is ``good''
for all integer $b\neq0 \bmod m$.
\end{lemma}

\comment{

The proof of the previous lemma doesn't give an algorithm for the
construction of a ``good'' set. But in \cite{an08} there is an
explicit approach based on ideas from \cite{aikps90}, which can be
used to construct a bigger, but still ``good'' set.

Fix $\epsilon>0$ and introduce the following notation:
$$P = \{p|\ p\mbox{ is prime and } (\log{m})^{1+\epsilon}/2<p\leq
(\log{m})^{1+\epsilon}\},$$
$$S = \{1, 2, \ldots, (\log{m})^{1+2\epsilon}\},$$
$$K = \{s\cdot p^{-1}|\ s\in S, p\in P\},$$
where $p^{-1}$ is the inverse modulo $m$.

It is obvious that $|K| = O(\log^{2+3\epsilon}m)$. It turns out
\cite{aikps90} that for each $g\in\{1, 2$, \ldots, $m-1\}$
$$\frac{1}{|K|}\left|\sum\limits_{j=1}^t e^{\frac{2\pi k_j g}{m}i}\right|\leq(\log{m})^{-\epsilon}.$$

Taking the real part of the previous inequality we induce that
$$\frac{1}{t}\left|\sum\limits_{j=1}^t \cos{\frac{2\pi k_j g}{m}}\right|\leq(\log{m})^{-\epsilon}$$
for each $g\in\{1, 2, \ldots, m-1\}$.

Thus, the set $K$ is ``good'' for all $g\neq 0 \bmod{m}$.

}

We use this result for our fingerprinting technique choosing  the
set $K=\{k_{1}, \dots, k_{t}\}$ which is ``good'' for all
$b=g(\sigma)\neq0$. That is, it allows to distinguish those inputs
whose image is 0 modulo $m$ from the others.

%%%%%%%%%%%%%%%%%%%%%%%%%%%%%%%%%%%%%%%%%%%%%%%%%%%%%%

\subsection{Boolean Functions Computable via Fingerprinting Method}
\comment{ It is obvious that the complexity of computing some
Boolean function $f$ via the fingerprinting technique propagates to
the creating of its fingerprint. Now the question is what types of
the mapping $g$ allow effective computation of the corresponding
fingerprint.

Generally, there are many polynomials for the same function. But not
all of them are suitable for our method. }

Let $f(x_1,\ldots, x_n)$ be a Boolean function and $g$ be its
characteristic polynomial. The following theorem holds.

\begin{theorem}\label{LinearPolynomialComputation}
Let $\epsilon\in(0,1)$. If $g$ is a linear polynomial over
$\mathbb{Z}_m$, i.e. $g=c_1x_1+\ldots c_nx_n+c_0$, then $f$ can be
computed with one-sided error $\epsilon$ by a quantum OBDD of width
$O\left(\frac{\log m}{\epsilon}\right)$.
\end{theorem}
\begin{proof}
Here is the algorithm in the circuit notation:
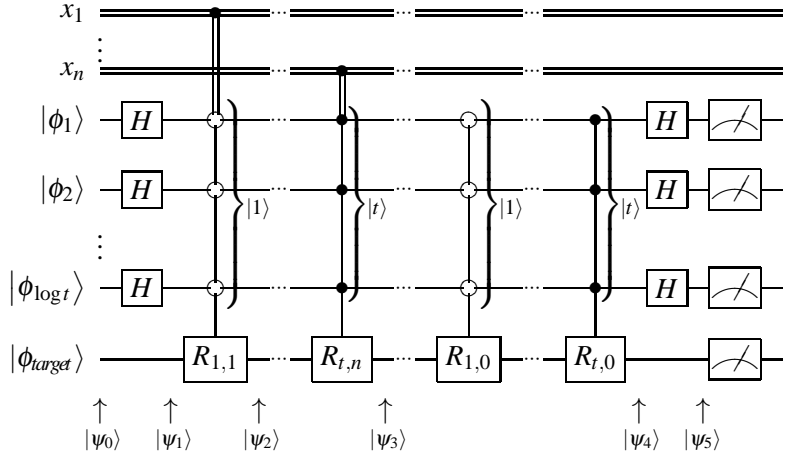
\[\quad\quad\quad
\Qcircuit @C=0.75em @R=1.0em {
\lstick{x_1} & \cw & \control\cw\cwx[3] & \cw & ~_{\cdots}\quad & \cw & \cw & ~_{\cdots}\quad & \cw & \cw & ~_{\cdots}\quad & \cw & \cw & \cw & \cw\\
\vdots \\
\lstick{x_n} & \cw & \cw & \cw & ~_{\cdots}\quad & \control\cw\cwx[1] & \cw & ~_{\cdots}\quad & \cw & \cw & ~_{\cdots}\quad & \cw & \cw & \cw & \cw\\
\lstick{\ket{\phi_1}} & \gate{H} & \ctrlo{1} & \qw  & ~_{\cdots}\quad & \ctrl{1} & \qw & ~_{\cdots}\quad & \ctrlo{1} & \qw  & ~_{\cdots}\quad & \ctrl{1} & \gate{H} & \meter & \qw\\
\lstick{\ket{\phi_2}} & \gate{H} & \ctrlo{2} & \qw  & ~_{\cdots}\quad & \ctrl{2} &  \qw & ~_{\cdots}\quad & \ctrlo{2} & \qw  & ~_{\cdots}\quad & \ctrl{2} & \gate{H} & \meter & \qw\\
\vdots &&\ustick{\quad\quad\quad^{\ket{1}}} &&& \ustick{\quad\quad~~^{\ket{t}}}&&& \ustick{\quad\quad\quad^{\ket{1}}}&&& \ustick{\quad\quad~~^{\ket{t}}}\gategroup{4}{3}{7}{3}{1em}{\}}\gategroup{4}{6}{7}{6}{1em}{\}}\gategroup{4}{9}{7}{9}{1em}{\}}\gategroup{4}{12}{7}{12}{1em}{\}} \\
\lstick{\ket{\phi_{\, \log t}}} & \gate{H} & \ctrlo{1} & \qw & ~_{\cdots}\quad & \ctrl{1}& \qw & ~_{\cdots}\quad & \ctrlo{1} & \qw  & ~_{\cdots}\quad & \ctrl{1} & \gate{H} & \meter & \qw\\
\lstick{\ket{\phi_{t\!a\!r\!g\!e\!t}}}& \qw & \gate{R_{1,1}} & \qw & ~_{\cdots}\quad & \gate{R_{t,n}} & \qw & ~_{\cdots}\quad & \gate{R_{1,0}} & \qw & ~_{\cdots}\quad & \gate{R_{t,0}} & \qw & \meter & \qw\\
\uparrow & \quad\quad\uparrow &\quad\quad\quad\uparrow&&& \quad\quad\quad\uparrow &&&&& & \quad\quad\quad\uparrow & \quad\quad~\uparrow\\
~_{\ket{\psi_0}} & \quad\quad~_{\ket{\psi_1}} &\quad\quad\quad~_{\ket{\psi_2}}&&& \quad\quad\quad~_{\ket{\psi_3}} &&&&& & \quad\quad\quad~_{\ket{\psi_4}} & \quad\quad~~_{\ket{\psi_5}}\\
}
\]

Initially qubits $\ket{\phi_1}\otimes\ket{\phi_2} \otimes\dots \otimes\ket{\phi_{\log
t}} \otimes \ket{\phi_{t\!a\!r\!g\!e\!t}}$ are in the state
$\ket{\psi_0}=\ket{0}^{\otimes\log{t}}\ket{0}$. For $i\in\{1,\dots, t\}$,
$j\in\{0,\dots, n\}$ we define rotations $R_{i,j}$ as
$$R_{i,j}=R_{\hat{y}}\left(\frac{4\pi k_i c_j}{m}\right),$$
where $c_j$ are the coefficients of the linear polynomial for $f$ and the set of
parameters $K=\{k_{1}, \dots, k_{t}\}$ is ``good'' according to the Definition
\ref{good-set} with $t=\GoodSetSize{\cdot m}$.

Let $\sigma=\sigma_1\ldots\sigma_{n} \in \{0,1\}^{n}$ be an input string.

The first layer of Hadamard operators transforms the state $\ket{\psi_0}$ into
$$\ket{\psi_1} = \frac{1}{\sqrt{t}}\sum\limits_{i=1}^t\ket{i}\ket{0}.$$

Next, upon input symbol $0$ identity transformation $I$ is applied. But if the value of
$x_j$ is $1$, then the state of the last qubit is transformed by the operator $R_{i,j}$,
rotating it by the angle proportional to $c_j$. Moreover, the rotation is done in each
of $t$ subspaces with the corresponding amplitude $1/\sqrt{t}$. Such a parallelism is
implemented by the controlled operators $C_i(R_{i,j})$, which transform the states
$\ket{i}\ket{\cdot}$ into $\ket{i}R_{i,j}\ket{\cdot}$, and leave others unchanged. For
instance, having read the input symbol $x_1=1$, the system would evolve into state
$$\begin{array}{rcl}
\ket{\psi_2} & = & \frac{1}{\sqrt{t}}\sum\limits_{i=1}^tC_i(R_{i,1})\ket{i}\ket{0} =
\frac{1}{\sqrt{t}}\sum\limits_{i=1}^t\ket{i}R_{i,1}\ket{0}\\
& = & \frac{1}{\sqrt{t}}\sum\limits_{i=1}^t\ket{i}\left(\cos\frac{2\pi k_i
c_1}{m}\ket{0} + \sin\frac{2\pi k_i c_1}{m}\ket{1}\right)
\end{array}.$$

Thus, after having read the input $\sigma$ the amplitudes would ``collect'' the sum
$\sum_{j=1}^nc_j\sigma_j$
$$
\begin{array}{rcl}
\ket{\psi_3} & = & \frac{1}{\sqrt{t}}\sum\limits_{i=1}^t\ket{i}\left(\cos\frac{2\pi k_i
\sum_{j=1}^nc_j\sigma_j}{m}\ket{0} + \sin\frac{2\pi k_i
\sum_{j=1}^nc_j\sigma_j}{m}\ket{1}\right)
\end{array}.
$$

At the next step we perform the rotations by the angle $\frac{4\pi k_i c_0}{m}$ about
the $\hat{y}$ axis of the Bloch sphere for each $i\in\{1,\dots,t\}$. Therefore, the
state of the system would be
$$\begin{array}{rcl} \ket{\psi_4} & = &
\frac{1}{\sqrt{t}}\sum\limits_{i=1}^t\ket{i}\left(\cos\frac{2\pi k_i
g(\sigma)}{m}\ket{0}+\sin\frac{2\pi k_i g(\sigma)}{m}\ket{1}\right).
\end{array}$$

Applying $H^{\otimes\log{t}}\otimes I$ we obtain the state
$$\begin{array}{rcl}
\ket{\psi_5} & = & \left(\frac{1}{t}\sum\limits_{i=1}^t\cos\frac{2\pi k_i
g(\sigma)}{m}\right)\ket{0}^{\otimes\log{t}}\ket{0}+\\
&& + \gamma\ket{0}^{\otimes\log{t}}\ket{1} +
\sum\limits_{i=2}^{t}\ket{i}\left(\alpha_i\ket{0} + \beta_i\ket{1}\right),
\end{array}$$
where $\gamma$, $\alpha_i$, and $\beta_i$ are some unimportant amplitudes.

The input $\sigma$ is accepted if the measurement outcome is
$\ket{0}^{\otimes\log{t}}\ket{0}$. Clearly, the accepting probability is
\[ Pr_{accept}(\sigma) = \frac{1}{t^2}\left(\sum\limits_{i=1}^t\cos\frac{
  2\pi k_i g(\sigma)}{2^{n}}\right)^2.
\]

If $f(\sigma)=1$ then $g(\sigma)=0$ and the program accepts $\sigma$ with probability
$1$. Otherwise, the choice of the set $K=\{k_1,\dots,k_t\}$ guarantees that
$$Pr_{accept}(\sigma) = \frac{1}{t^2}\left(\sum\limits_{i=1}^t\cos\frac{
  2\pi k_i g(\sigma)}{2^{n}}\right)^2<\epsilon.$$

Thus, $f$ can be computed by a $q$-qubit quantum OBDD, where $q=\log{2t}=O(\log\log m)$.
The width of the program is $2^q=O(\log m)$.
\end{proof}

The following functions have the aforementioned linear polynomials
and thus are effectively computed via the fingerprinting technique.

\paragraph{$MOD_m$} The function $MOD_m$ tests whether the number of $1$'s in the input
is $0$ modulo $m$. The linear polynomial over $\mathbb{Z}_m$ for
this function is
$$\sum\limits_{i=1}^nx_i.$$
The lower bound for the width of deterministic OBDDs computing this
function is $\Omega(m)$ \cite{weg00}. Thus, our method provides an
exponential advantage of quantum OBDD over any deterministic one.

\comment{

\paragraph{$MOD'_m$} This function is the same as $MOD_m$, but the input
is treated as binary number. Thus, the linear polynomial is
$$\sum\limits_{i=1}^nx_i2^{i-1}.$$

}

\paragraph{$EQ_n$} The function $EQ_n$, which tests the equality of two
$n$-bit binary strings, has the following polynomial over
$\mathbb{Z}_{2^n}$
$$\sum\limits_{i=1}^nx_i2^{i-1}- \sum\limits_{i=1}^ny_i2^{i-1}.$$%\bmod{2^n}.$$
%The lower bound for the width of deterministic OBDDs computing this
%function is $\Omega(m)$.

\paragraph{$Palindrome_n(x_1,\ldots,x_n)$} This function tests the symmetry of the
input, i.e. whether $x_1 x_2\ldots x_{\lfloor n/2\rfloor}$ = $x_n
x_{n-1}\ldots x_{\lceil n/2\rceil+1}$ or not. The polynomial over
$\mathbb{Z}_{2^{\lfloor n/2\rfloor}}$ is
$$\sum\limits_{i=1}^{\lfloor n/2\rfloor}x_i2^{i-1}- \sum\limits_{i=\lceil n/2\rceil}^nx_i2^{n-i}.$$%\bmod{2^{\lfloor n/2\rfloor}}$$

\comment{

\paragraph{$Period_n^s(x_{0},\ldots, x_{n-1})$} $= 1$
iff $x_i=x_{i+s\bmod{n}}$ for all $i\in\{0,\ldots,n-1\}$. The
polynomial over $\mathbb{Z}_{2^n}$ is
$$\sum\limits_{i=0}^{n-1}x_i\left(2^{i}-2^{i-s\bmod{n}}\right).$$%\bmod{2^n}$$

\paragraph{$Semi-Simon_n^s(x_{0},\ldots, x_{n-1})$} $= 1$
iff $x_i=x_{i\oplus s}$ for all $i\in\{0,\ldots,n-1\}$. The
polynomial over $\mathbb{Z}_{2^n}$ is
$$\sum\limits_{i=0}^{n-1}x_i\left(2^{i}-2^{i\oplus
s}\right).$$%\bmod{2^n}$$

}

\paragraph{$PERM_n$} The \emph{Permutation Matrix} test function ($PERM_n$) is defined on
$n^2$ variables $x_{i j}$ ($1\leq i,j\leq n$). It tests whether the
input matrix contains exactly one 1 in each row and each column.
Here is a polynomial over $\mathbb{Z}_{(n+1)^{2n}}$
$$\sum\limits_{i=1}^n\sum\limits_{j=1}^nx_{ij}\left((n+1)^{i-1}+(n+1)^{n+j-1}\right)
            - \sum\limits_{i=1}^{2n}(n+1)^{i-1}.$$%\bmod{(n+1)^{2n}}.$$

Note, that this function cannot be effectively computed by a
deterministic OBDD -- the lower bound is $\Omega(2^nn^{-5/2})$
regardless of the variable ordering \cite{weg00}. The width of the
best known probabilistic OBDD, computing this function with
one-sided error, is $O(n^4\log{n})$ \cite{weg00}. Our algorithm has
the width $O(n\log n)$. Since the lower bound $\Omega(n-\log{n})$
follows from Theorem \ref{quant-det-width}, our algorithm is almost
optimal.

The following table provides the comparison of the width of quantum
and deterministic OBDDs for the aforementioned functions.

\begin{center}
\begin{tabular}{|l|l|l|}
  \hline
  % after \\: \hline or \cline{col1-col2} \cline{col3-col4} ...
  & OBDD & QOBDD \\
  \hline
  $MOD_m$ & $\Omega(m)$ & $O(\log{m})$ \\
  \hline
%  $MOD'_m$ & $\Omega(m)$ & $O(\log{m})$ \\
%  \hline
  $EQ_n$ & $2^{\Omega(n)}$ & $O(n)$ \\
  \hline
  $Palindrome_n$ & $2^{\Omega(n)}$ & $O(n)$ \\
  \hline
  $PERM_n$ & ${\Omega(2^nn^{-5/2})}$ & $O(n\log{n})$ \\
  \hline
%  $HSP_{G,K}$ & $...$ & $O(|G|\log G:K)$ \\
%  \hline
\end{tabular}
\end{center}

\section{Generalized Approach}

The fingerprinting technique described in the previous section
allows us to test a single property of the input encoded by a
characteristic polynomial. Using the same ideas we can test the
conjunction of several conditions encoded by a group of
characteristic polynomials which we call a \emph{characteristic} of
a function.

\begin{definition}
We call a set $\chi_f^m$ of polynomials over $\mathbb{Z}_m$ a
\emph{characteristic} of a Boolean function $f$ if for all
polynomials $g\in\chi_f^m$ and all $\sigma\in\{0,1\}^n$ it holds
that $g(\sigma)=0$ iff $\sigma\in f^{-1}(1)$.
\end{definition}

We say that a characteristic is \emph{linear} if all of its
polynomials are linear.

From Lemma \ref{Existence-Of-A-Characteristic-Polynomial} it follows
that for each Boolean function there is always a characteristic
consisting of a single characteristic polynomial.

\comment{

\begin{lemma}
For any Boolean function $f$ there exists a linear characteristic
$\chi_f^{n+1}$ of size $O(|f^{-1}(1)|)$.
\end{lemma}

\begin{proof}
One way to construct such characteristic over ${\mathbb Z}_{n+1}$ is
transforming a DNF for $f$.

Let $K_1\vee\ldots\vee K_l$ be a DNF for $f$ and let $g_i$ be a sum
of variables from $K_i$ (negations $\overline{x}_j$ are replaced by
$1-x_j$) minus $l_i$:
$$K_i=x_{i_1}^{\delta_{i_1}}\wedge\ldots \wedge x_{i_{l_i}}^{\delta_{i_{l_i}}}\Longrightarrow
g_i=\sum\limits_{j=1}^{l_i}\left(\delta_jx_j+(1-\delta_j)(1-x_j)\right)-l_i.$$
That is, $g_i$ ``tests'' whether $K_i=1$ or not by counting the
number of ones in the product.

If $f=1$, then at least one of $K_i$ equals $1$ and the
corresponding $g_i$ equals $0$. Then the set of all $g_i$ is a
characteristic for $f$ with $m=n+1$.
\end{proof}

}

Now we can generalize the Fingerprinting technique from section
\ref{fqbp}.

\paragraph{Generalized Fingerprinting technique} For a Boolean function $f$ we
choose an error rate $\epsilon>0$ and pick a characteristic
$\chi_f^m=\{g_1,\ldots, g_l\}$. Then for arbitrary binary string
$\sigma=\sigma_1\ldots \sigma_n$ we create its fingerprint
$\ket{h_\sigma}$ composing $t\cdot l$ ($t=\GoodSetSize{m}$) single
qubit fingerprints $\ket{h^i_\sigma(j)}$:
$$\begin{array}{rcl}
\ket{h^i_\sigma(j)} & = & \cos\frac{\pi k_i
g_j(\sigma)}{m}\ket{0}+\sin\frac{\pi k_i g_j(\sigma)}{m}\ket{1}\\
\ket{h_\sigma} & = &
\frac{1}{\sqrt{t}}\sum\limits_{i=1}^t\ket{i}\ket{h^i_\sigma(1)}\ket{h^i_\sigma(2)}\ldots\ket{h^i_\sigma(l)}
\end{array}$$
%That is, the last qubit is rotated by $t$ different angles about the
%$\hat{y}$ axis of the Bloch sphere.

\begin{theorem}\label{LinearCharacteristicComputation}
If $\chi_f^m$ is a linear characteristic then $f$ can be computed by
a quantum OBDD of width $O(2^{|\chi_f^m|}\log m)$.
\end{theorem}
\begin{proof}
Here is the sketch of the algorithm:
\begin{enumerate}
\item Upon the input $\sigma=\sigma_1\ldots\sigma_n$ we create the
fingerprint $\ket{h_\sigma}$.
\item We measure $\ket{h_\sigma}$ in the standard computational
basis and accept the input if the outcome of the last $l$ qubits
is the all-zero state. Thus, the probability of accepting
$\sigma$ is
$$Pr_{accept}(\sigma) = \frac{1}{t}\sum\limits_{i=1}^t\cos^2\frac{
  \pi k_i g_1(\sigma)}{m}\cdots \cos^2\frac{
  \pi k_i g_l(\sigma)}{m}.$$

If $f(\sigma)=1$ then all of $g_i(\sigma)=0$ and we will always
accept.

If $f(\sigma)=0$ then there is at least one such $j$ that
$g_j(\sigma)\neq0$ and the choice of the ``good'' set $K$
guarantees that the probability of the erroneously accepting is
bounded by
$$\begin{array}{rcl}
Pr_{accept}(\sigma) & = & \frac{1}{t}\sum\limits_{i=1}^t\cos^2\frac{
  \pi k_i g_1(\sigma)}{m}\cdots \cos^2\frac{\pi k_i g_l(\sigma)}{m}\\
& \leq & \frac{1}{t}\sum\limits_{i=1}^t\cos^2\frac{\pi k_i
g_j(\sigma)}{m} =
  \frac{1}{t}\sum\limits_{i=1}^t\frac{1}{2}\left(1+\cos\frac{
  2\pi k_i g_j(\sigma)}{m}\right) \\
& = & \frac{1}{2}+\frac{1}{2t}\sum\limits_{i=1}^t\cos\frac{2\pi k_i g_j(\sigma)}{m}\\
& \leq & \frac{1}{2} + \frac{\sqrt{\epsilon}}{2}.
\end{array}$$
\end{enumerate}

The number of qubits used by this QBP is $q=O(\log\log m + l)$,
$l=|\chi_f^m|$. Therefore, the width of the program is
$2^q=O(2^{|\chi_f^m|}\log m)$.

\end{proof}

The generalized approach can be used to construct an effective
quantum OBDD for the Boolean variant of the {\em Hidden Subgroup
Problem}.

\subsection{The upper bound for Hidden Subgroup Function}

This problem was first defined and considered in \cite{khasdis},
where the following Boolean variant of the \emph{Hidden Subgroup
Problem} was defined.

% The
%proof of the theorem in this section follows somewhat different
%presentation from \cite{khasdisrus}. %In this paper we give a shorter
%and more illustrative proof of the result via circuit presentation,
%the approach first applied in \cite{av08}.

%In order to investigate Quantum Branching Program complexity of the
%\emph{Hidden Subgroup Problem}, we define a function.

\begin{definition} \label{hspdfn} Let $K$ be a normal subgroup of a finite group $G$. Let $X$ be
a finite set. For a sequence $\chi\in X^{|G|}$ let $\sigma=bin(\chi)$ be its
representation in binary. If $\sigma$ encodes no correct sequence
$\chi=\chi_1\ldots\chi_{|G|}$, then \emph{Hidden Subgroup} function of $\sigma$ is set
to be zero, otherwise:
$$
\hsp{\sigma}=\left\{
\begin{array}{ll}
1, & \text{if }~~ \forall a\in G~ \forall~ i,j \in aK~ (\chi_i=\chi_j)\\
& \text {and }~~ \forall~ a,b\in G~ \forall~ i \in aK~ \forall~ j \in bK~
 (aK\neq bK\Rightarrow \chi_{i}\neq\chi_{j});\\
0, & \text{otherwise.}
\end{array}
\right. $$
\end{definition}

Let $f$ be the function encoded by the input sequence. We want to
know if a function $f:G\rightarrow X$ ``hides'' the subgroup $K$ in
the group $G$. Our program receives $G$ and $K$ as
\emph{parameters}, and function $f$ as an \emph{input string}
containing values of $f$ it takes on $G$. The values are arranged in
lexicographical order. See Definition \ref{hspdfn}.

We make two assumptions. First, we assume that the set $X$ contains
exactly $(G:K)$ elements. Indeed, having read the function $f$,
encoded in the input sequence $\sigma$, we have $X$ to be the set of
all different values that $f$ takes. Obviously, if $|X|$ is less or
greater than $(G:K)$, then $\hsp{\sigma}=0$. The second assumption,
is that we replace all values of $f$ by numbers from $1$ through
$(G:K)$. Thus, $\hsp{x_1,\ldots,x_n}$ is a Boolean function of
$n=|G|\lceil\log{G:K}\rceil$ variables. In these two assumptions the
following theorem holds.

\begin{theorem}\label{HSP-Upper-Bound} Function \hsp{x} can be computed with one-sided error
by a quantum OBDD of width $O(n)$.
\end{theorem}

\begin{proof}
First we shall prove the following lemma.

\begin{lemma}\label{alglem}
In order to correctly compute $\hsp{x}$ it is enough to perform
following calculations.
\begin{enumerate}
\item For every coset we check equalities for all input sequence values that have
indices from this coset;
\item From every coset we choose a representative, and check if the sum of values of $f$
on all the representatives equals to the following value
$$
S=\sum_{i=1}^{G:K}i=\frac{(G:K)((G:K)+1)}{2}.
$$
\end{enumerate}
\end{lemma}
\begin{proof}
One direction is straightforward. The other direction is also not
difficult. Suppose we have the two conditions of the lemma
satisfied. Let $aK$ and $bK$ be two different cosets with elements
$d\in aK$ and  $c\in bK$, such that $\sigma_{d}=\sigma_{c}$. We fix
$c\in bK$. There are two cases possible:
\begin{enumerate}
\item For all $d\in aK (\sigma_{d}=\sigma_{c})$;
\item There exists $d' \in aK (\sigma_{d}\neq\sigma_{c})$.
\end{enumerate}
Apparently in the first case we indeed could choose any of the
elements of a coset to check inequalities. In the second case the
first condition of the lemma would fail. The reasoning for $bK$ is
analogous.

When the values of $f$ are different on different cosets, obviously,
the sum of these values is the sum of numbers from $1$ through
$G:K$. Therefore, $\hsp{\sigma}=1$ iff both conditions of the lemma
are satisfied.
\end{proof}

According to the previous lemma, $\hsp{x}$ has a characteristic
consisting of two polynomials over $\mathbb{Z}_{2^n}$, checking
conditions of the lemma. We shall construct them explicitly to show
they are linear.

We shall adopt another indexation of $\chi$ when convenient:
$\chi_{a,q}$ is a value of $f$ on the $q$-th element of the coset
$aK$.

Therefore, for a binary input symbol $x_j$ we define
\begin{itemize}
\item $a=a(j)$ for the number of the corresponding coset;
\item $q=q(j)$ for the number of the corresponding element of the coset $a$;
\item $r=r(j)$ for the number of bit in the binary representation of $\chi_{a,q}$
\end{itemize}
and start indexation from 0. Thus $a\in\{0,\ldots,(G:K)-1\},
q\in\{0,\ldots,|aK|-1\}$.

In this notation the polynomials are:
\begin{enumerate}
\item $g_1(x)=\sum_a\sum_{q}2^{(|K|a+q)\lceil\log{G:K}\rceil}(\chi_{a,{q}}-\chi_{a,{q-1\bmod{|K|}}})$.
Thus, $g_1(x)=0$ iff for every coset $a$ function $f$ maps all
the elements of $a$ onto the same element of $X$.
\item $g_2(x)=\left(\sum_{j=1}^{(G:K)}\chi_{i_j}\right) - S$, where $\chi_{i_j}$ is
the representative chosen from the $j$-th coset. Therefore,
$g_2(x)$ checks whether the images of elements from different
cosets are distinct.
\end{enumerate}

By the generalized fingerprinting technique we can construct quantum
OBDD of width $O(n)$, computing $\hsp{x}$ with one-sided error.
\end{proof}

\bibliographystyle{eptcs}

%\appendix
%\comment{

\newpage
\begin{appendix}

\section{Proof of Lemma \ref{existence-of-a-good-set}}
\begin{proof}
Using Azuma's inequality (see, e.g., \cite{mr95}) we prove that a
random choice of the set $K$ is ``good'' with positive probability .

Let $1\leq g\leq m-1$ and let $K$ be the set of $t$ parameters
selected uniformly at random from $\{0,\dots, m-1\}$.

We define random variables $X_i=\cos\frac{2\pi k_i g}{m}$ and
$Y_k=\sum_{i=1}^k X_i$. We want to prove that Azuma's inequality is
applicable to the sequence $Y_0=0$, $Y_1$, $Y_2$, $Y_3$, \ldots,
i.e. it is a martingale with bounded differences. First, we need to
prove that $E[Y_k]<\infty$.

From the definition of $X_i$ it follows that
$$E[X_i]=\frac{1}{m}\sum\limits_{j=0}^{m-1}\cos\frac{2\pi j g}{m}$$

Consider the following weighted sum of $m$th roots of unity
$$\frac{1}{m}\sum\limits_{j=0}^{m-1}\exp\left(\frac{2\pi j g
}{m}i\right)=\frac{1}{m}\cdot\frac{\exp(2\pi i g m /m)-1}{\exp(2\pi
i g /m)-1} = 0,$$ since $g$ is not a multiple of $m$.

$E[X_i]$ is exactly the real part of the previous sum and thus is
equal to $0$.

Consequently, $E[Y_k]=\sum_{i=1}^k E[X_i] = 0 < \infty$.

Second, we need to show that the conditional expected value of the
next observation, given all the past observations, is equal to the
last observation.
$$E[Y_{k+1}|Y_1,\ldots,Y_k] = \frac{1}{m}\sum\limits_{j=0}^{m-1}\left(Y_k +
\cos\frac{2\pi j g}{m}\right)=Y_k +
\frac{1}{m}\sum\limits_{j=0}^{m-1}\cos\frac{2\pi j g}{m} = Y_k$$

Since $|Y_{k+1}-Y_k|=|X_{k+1}|\leq1$ for $k\geq0$ we apply Azuma's
inequality to obtain
$$Pr(|Y_t-Y_0|\geq\lambda)=Pr\left(|\sum\limits_{i=1}^tX_i|\geq\lambda\right)\leq
2\exp\left(-\frac{\lambda^2}{2t}\right)$$

Therefore, we induce that the probability of $K$ being not ``good''
for $1\leq g\leq m-1$ is at most
$$Pr\left(|\sum\limits_{i=1}^t X_i|\geq\sqrt{\epsilon}t\right)\leq
2\exp\left(-\frac{\epsilon t}{2}\right)\leq\frac{1}{m}$$ for
$t=\lceil(2/\epsilon)\ln{2m}\rceil$.

Hence the probability that constructed set is not ``good'' for at
least one $1\leq g\leq m-1$ is at most $(m-1)/m<1$. Therefore, there
exists a set which is ``good'' for all $1\leq g\leq m-1$. This set
will also be ``good'' for all $g\neq0 \bmod m$ because
$\cos\frac{2\pi k (g+jm)}{m}=\cos\frac{2\pi k g}{m}$.
\end{proof}

\end{appendix}

%}
%%%%%%%%%%%%%%%%%%%%%%%%%%%%%%%%%%%%%%%%%%%%%%%%%%%%%%

\end{document}